\documentclass[11pt]{article}

\usepackage{amssymb,amsfonts,amsthm,amsmath,amssymb,amscd,amstext,epsfig}

\def\ben{\begin{equation}}
\def\een{\end{equation}}

\let\a=\alpha \let\b=\beta  \let\d=\delta 
  \let\q=\theta 
\let\l=\lambda  \let\n=\nu

   \let\Q=\Theta 
 \let\P=\Phi

\let\pa=\partial
\def\be{\begin{equation}}
\def\ee{\end{equation}}
\def\ba{\begin{array}}
\def\ea{\end{array}}

\def\dalemb#1#2{{\vbox{\hrule height .#2pt
        \hbox{\vrule width.#2pt height#1pt \kern#1pt
                \vrule width.#2pt}
        \hrule height.#2pt}}}

\newcommand{\bea}{\begin{eqnarray}}
\newcommand{\eea}{\end{eqnarray}}

\newcommand{\Tr}{{\rm Tr} }
\def\ep{{\epsilon}}

\def\R{{{\mathbb R}}}

\def\zb{{\bar z}}

\begin{document}
\begin{flushright}
NSF-KITP-07-193 \\
NI-07076\\
arXiv:0711.3026 [hep-th]
\end{flushright}

\begin{center}
\vspace{1cm} { \LARGE {\bf Strings on conifolds from strong coupling dynamics: quantitative results}}

\vspace{1.1cm}

David E. Berenstein$^\sharp$ and Sean A. Hartnoll$^\flat$

\vspace{0.8cm}

{\it $^\sharp$ Department of Physics, University of California\\
     Santa Barbara, CA 93106-9530, USA \\

\vspace{0.3cm}

$^\sharp$ Isaac Newton Institute for Mathematical Sciences\\
Cambridge CB3 0EH, UK }

\vspace{0.5cm}

{\it $^\flat$ KITP, University of California\\
     Santa Barbara, CA 93106-4030, USA }

\vspace{0.8cm}

{\tt dberens@physics.ucsb.edu, hartnoll@kitp.ucsb.edu} \\

\vspace{1.2cm}

\end{center}

\begin{abstract}
\noindent
Three quantitative features of string theory on $AdS_5 \times
X_5$, for any (quasi)regular Sasaki-Einstein $X_5$, are recovered
exactly from an expansion of field theory at strong coupling
around configurations in the moduli space of vacua. These
configurations can be thought of as a generalized matrix model of
(local) commuting matrices. First, we reproduce the spectrum of
scalar Kaluza-Klein modes on $X_5$. Secondly, we recover the
precise spectrum of BMN string states, including a nontrivial
dependence on the volume of $X_5$. Finally, we show how the radial
direction in global $AdS_5$ emerges universally in these theories
by exhibiting states dual to AdS giant gravitons.

\end{abstract}

\pagebreak
\setcounter{page}{1}

\section{Introduction}

The AdS/CFT correspondence \cite{Maldacena:1997re} provides,
in principle, a nonperturbative approach to quantum gravity in
asymptotically Anti-de Sitter space. A traditionally thorny issue
in quantum gravity is the emergence of spacetime and gravitons in
a semiclassical limit. In AdS/CFT, addressing this question requires us
to directly tackle the dual strongly coupled conformal field theory,
in the large $N$ limit. This is a different sort to problem to much
work that has been done in AdS/CFT,
in which protected quantities, or integrable sectors of
the theory, are computed at weak and strong coupling and compared
directly.

A program aimed at understanding the emergence of semiclassical
quantum gravity from field theory was initiated in \cite{Berenstein:2005aa}.
The starting point is a guess concerning the effective, semiclassical,
degrees of freedom which characterize the ground state and
dominate the low energy physics of the strongly coupled theory,
together with a proposal for their dynamics.
We will review aspects of this proposal below. Using this effective
low energy theory, various non protected quantities were computed
and successfully compared with the dual string theory
\cite{Berenstein:2005jq, Berenstein:2007zf, Berenstein:2007wz}.
Furthermore, the proposal was extended from the original case
of ${\mathcal{N}}=4$ Super Yang-Mills theory to orbifolds of this
theory in \cite{Berenstein:2005ek, Berenstein:2006yy}.

It was recently argued \cite{Berenstein:2007wi} that the original proposal,
which was for the ${\mathcal{N}}=4$
theory, can be generalized to a large class of
conformal field theories with only ${\mathcal{N}}=1$
supersymmetry. In particular to the theories arising on
$N$ D3 branes at the tip of a Calabi-Yau cone. This is a substantial
generalization, as there are many such theories. In fact, these theories
are in one to one correspondence with the space of five dimensional
Sasaki-Einstein metrics \cite{Morrison:1998cs}.

In this paper we will use and extend the recent proposal
\cite{Berenstein:2007wi} to derive various quantities in the
strongly coupled ${\mathcal{N}}=1$ theories. These will be non-BPS
quantities and they will reproduce in detail the dual, spacetime, AdS gravity
results. We start in sections 3 and 4 by obtaining the ground
state of the effective theory and showing that it describes the
emergence of the dual Sasaki-Einstein geometry. In sections 5, 6
and 7 we study fluctuations about the ground state, reproducing
the spacetime spectrum of scalar Kaluza-Klein harmonics and of BMN
string states. Finally, in section 8 we show, by considering
excitations dual to giant gravitons, that the radial direction of
$AdS_5$ emerges universally, i.e. orthogonally to and
independently of the internal Sasaki-Einstein manifold. In the
concluding discussion we emphasize the many computations that
remain to be done in order to flesh out this framework in detail.

\section{Summary of the computational framework}

The upshot of the detailed arguments in \cite{Berenstein:2007wi}
will now be summarized, together with some new statements which we
will expand upon in later sections. The objective is to describe
the low energy physics of the strongly coupled superconformal
theories arising on $N$ D3 branes at the tip of a Calabi-Yau cone.
The field theory is on a spatial $S^3$ and hence dual to global
$AdS_5$ space.

\begin{itemize}

\item The degrees of freedom which dominate the large $N$ low
energy dynamics are configurations of scalar fields that explore
the moduli space of vacua of the field theory. The scalar fields
are uniform on the $S^3$, that is, only the s-wave modes are
excited. Locally on the moduli space this is very similar to ${
\cal N}=4$ SYM, where the configurations
are given by six commuting $N\times N$ matrices. Thus in
the first instance we have integrated out the higher harmonics on
the spatial $S^3$, all the gauge fields and fermions, and all the
(generalized) off diagonal modes.

\item The $N$ eigenvalues of these matrices, $\{x_i\}$, are valued on a
Calabi-Yau cone over a Sasaki-Einstein manifold $X_5$. This would
be the moduli space of the theory on $\R^3$. Placing the theory on
$S^3$ lifts the moduli space. Firstly because of the conformal
coupling mass term. Secondly because there is
an enhanced symmetry $U(1)^2 \to U(2)$ when two eigenvalues
coincide, the measure terms arising from this degeneration
induces a repulsion between eigenvalues. The competition
between these two effects is captured by the Hamiltonian
\be\label{eq:hamil}
H = \sum_i \left[- \frac{1}{2 \mu^2}
\nabla_i \cdot (\mu^2 \nabla_i) + K_i \right] \,.
\ee
This expression is a sum of single particle Hamiltonians, labeled
by the subscript $i$, except for the measure factor $\mu^2$ which
depends on the locations of all the eigenvalues. Here $K$ is the
K\"ahler potential of the Calabi-Yau.

\item The measure factor $\mu^2$ requires an inspired guess. In
the ${\mathcal{N}}=4$ theory we have access to a weakly coupled
regime in which the measure factor can be directly determined
\cite{Berenstein:2005aa}. This is also possible in the case of
orbifolded theories \cite{Berenstein:2005ek, Berenstein:2006yy}.
One can then use nonrenormalization theorems for the BPS sector of
the theory to argue that the expression remains valid at strong
coupling. The conjectured form we use here, generalizing a
property of the measure in the ${\mathcal{N}}=4$ case, is that
\be\label{eq:mes}
\mu^2 = e^{-\sum_{i \neq j} s_{ij}} \,,
\ee
where $s_{ij}$ is the Green's function of a sixth order
differential operator on the Calabi-Yau cone
\be
- \nabla^6 s(x,x') = 64 \pi^3 \delta^{(6)}(x,x') \,.
\ee
As we will discuss below, this expression has the virtue of
automatically localizing the large $N$ eigenvalue distribution on
a hypersurface in the Calabi-Yau cone and thus leading to an
emergent geometry.

\item Given the Hamiltonian (\ref{eq:hamil}), one can find the ground state. We do
this in section 3 below. The answer is simply
\be\label{eq:gnd}
\psi_0 = e^{- \sum_i K_i} \,.
\ee
In section 4 below we show that in this state in the large $N$
limit the eigenvalues form an $X_5$ at fixed radius $r$ in the
cone, which we compute. This is to be interpreted as the $X_5$ of
the dual geometry, which has emerged from the matrix quantum
mechanics.

\item Given the ground state (\ref{eq:gnd}), one can find the spectrum of low
lying excitations. There are three types of excitations to
consider. The first are those given by operators made
from the six matrices that appear in the matrix quantum mechanics.
The energies of these states are given by the spectrum of the
Hamiltonian (\ref{eq:hamil}). An important set of eigenstates,
that we will consider, are of the form
\be
\psi = \psi_0 \Tr f(x) \,,
\ee
where $f(x)$ is some function of the six matrices that has polynomial growth.

Secondly, there are excitations of the off diagonal modes of the
six matrices, which commute in the ground state. These require
additional input. It was argued in \cite{Berenstein:2007wi} that
the physics of off diagonal modes connecting nearby eigenvalues is
the same as that of the ${\mathcal{N}}=4$ theory, with an
effective, ${\mathcal{N}}=4$ coupling $g_{\text{eff.}}$. In
particular, this implies that the energy of the mode connecting
nearby eigenvalues $x_i$ and $x_j$ is
\cite{Berenstein:2005aa, Berenstein:2005jq}
\be
E_{ij} = \sqrt{1 + \frac{g_\text{eff.}^2}{2 \pi^2}
 | x_i - x_j |^2 }
\,,
\ee
where the distance is given by the metric on the Calabi-Yau cone.
We require moreover that $g^2_{\text{eff.}}N$ is large in the
sense of 't Hooft.

Thirdly, there are excitations of the fields that have been set to
zero in the quantum mechanics: the higher harmonics of fields on
$S^3$, the gauge fields and the fermions. These modes remain
largely unexplored, although see \cite{Berenstein:2007zf, Aharony:2007rj}.

\end{itemize}

With this framework, the strategy for computing quantities is as follows.
Firstly we
compute the ground state wavefunction of the Hamiltonian. We can
then compute the energies of excitations about the ground state.
These will not in general be BPS. We show that the spectrum of
various excitations matches that computed in supergravity and
string theory, providing evidence for the calculational recipe
just presented.

\section{The ground state wavefunction}

The eigenvalue dynamics takes place on the six dimensional cone
over a five dimensional compact manifold $X_5$
\be\label{eq:metric}
ds^2_6 = dr^2 + r^2 ds^2_{5} \,.
\ee
Denote the coordinates on the five dimensional manifold by $\q$.
As we have mentioned, an important ingredient for writing down the
Hamiltonian for these eigenvalues is the Green's function on the
cone satisfying
\be\label{eq:greens}
- \nabla^6_6 s(r,r',\q,\q') = - \left(\frac{1}{r^5}
\frac{d}{dr} r^5 \frac{d}{dr} + \frac{1}{r^2} \nabla^2_5 \right)^3
s(r,r',\q,\q') = 64 \pi^3 \delta^{(6)}(r,r',\q,\q') \,.
\ee
This Green's function appears in the measure that is necessary to
write the Hamiltonian as a differential operator. See equations
(\ref{eq:hamil}) and (\ref{eq:mes}) above. We will now motivate
the use of this Green's function.

In the case of ${\cal N}=4 $ SYM the measure arising in going to
an eigenvalue description can be calculated, and is given by a
generalized Vandermonde determinant
\be
\mu^2 = \prod_{i<j} |\vec x_i-\vec x_j|^2 \,,
\ee
where we use vector notation to indicate a point in $\R^6$ (which
is the cone over $S^5)$. We notice that this function is
factorized over pairs of eigenvalues, and that if we take (minus
one times) the logarithm of $|\vec x_i -\vec x_j|^2$, then it
satisfies the differential equation (\ref{eq:greens}).

A similar calculation was done for abelian orbifolds  by a group
$\Gamma$ of ${\cal N}=4 $ SYM, and the corresponding measure was
also factorized \cite{Berenstein:2005ek,Berenstein:2006yy}. It was found that
\be
\mu^2 = \prod_{i<j}\prod_{\gamma\in \Gamma} |\vec x_i - \gamma(\vec
x_j)|^2\,.
\ee
That is, in the logarithm of the measure, we need to take a sum
over images to obtain the correct answer. Because we are summing
over images, the measure function obtained this way naturally
satisfies the Green's function equation (\ref{eq:greens}) for $X_5
= S^5/\Gamma$.

These measures were calculated by doing a one loop calculation in
the gauge theory: the volume of the gauge orbit of the
configurations. If we take a general conformal field theory, the
field theories will usually not have a weak coupling description.
This is because the fundamental fields have large anomalous
dimensions. We need some substitute for the one loop calculation
that preserves the spirit of the problem.

In \cite{Berenstein:2007wi} it was argued that the measure in the
general case should also be pairwise factorized and that in the
limit of coinciding points, the degeneration should be identical
to the case of ${\cal N}=4 $ SYM. This was argued by an effective
field theory reasoning and is true exactly in the case of the
orbifold measure. This suggests that the singularity in the
logarithm of the measure should be reproduced for all cases.
Choosing the Green's function above guarantees this behavior. In
principle, there could be other choices.

In the theories that admitted a Gaussian fixed point, the origin
of the measure was the volume of a gauge orbit. One might have
anticipated that this is the correct property to generalize.
However, there can be many different theories in the UV that can
give rise to the same conformal fixed point. This observation is
due to Seiberg \cite{Seiberg:1994pq}, and the different theories
are related by Seiberg dualities. If we examine various examples
of these theories, we find that generically the dimension of the
manifold associated to a single gauge orbit changes between
different dual theories and this would indicate that the measure
factor changes its scaling properties. However, we expect that the
effective dynamics should not change at all. Considering that
theories at strong coupling could behave very differently than at
weak coupling, calculating the effective measure by just measuring
the volume of the gauge orbit is suspect. Instead, it seems more
natural that whatever the effective dynamics is, it should depend
only on properties of the moduli space of vacua, as these are
automatically invariant under Seiberg dualities. Solving a
differential equation on the moduli space has this property.

In the end, we have to make a guess. The one we have made,
equation (\ref{eq:mes}), seems the simplest guess for the measure
term that matches all known cases. Our final justification will
perforce come a posteriori, after we have shown that this measure
gives many desirable features, especially at the level of
calculability of various properties of the strong coupling
dynamics.

Given the Hamiltonian, we now want to find wave functions that
solve it. In particular, to determine the ground state of the
Hamiltonian, we will need to know how the Green's function scales
under $r,r' \to \a r, \a r'$. In appendix A we show that the
Green's function obeys a logarithmic scaling
\be\label{eq:scaling}
s(\a r,\a r',\q,\q') = s(r,r',\q,\q') - \frac{\pi^3 \log \a}{
\text{Vol}(X_5)} \,.
\ee
It is interesting to note that the appearance of a nontrivial
scaling is intimately tied up with the need to regularize the
Green's function. In this sense, the scaling symmetry might be
called `anomalous'.

We now assume that $X_5$ is a Sasaki-Einstein manifold (see e.g.
\cite{Morrison:1998cs, Gibbons:2002th, Martelli:2006yb}) so that
the six dimensional cone is Calabi-Yau. Let
$\{z^a_i,\zb^{\bar a}_i\}$ be the complex coordinates of the $i$th
eigenvalue on the cone, $a,
\bar a = 1..3$. Let $K$ be the K\"ahler potential of the
Calabi-Yau.

The conjectured Hamiltonian \cite{Berenstein:2007wi} is
\be\label{eq:hamiltonian}
H = \sum_i \left( - \frac{g^{a \bar b}(z_i, \zb_i)}{2 \mu^2}
\left[\nabla_{z_i^a} \left( \mu^2 \nabla_{\zb_i^{\bar b}} \right) + \nabla_{\zb_i^{\bar b}} \left( \mu^2 \nabla_{z_i^a} \right) \right]+ K(z_i, \zb_i)
\right)
\,,
\ee
where the measure factor is
\be\label{eq:measure}
\mu^2 = e^{- \sum_{i \neq j} s_{ij}} \,.
\ee
Here $s_{ij}= s(z_i,\zb_i,z_j,\zb_j)$ is the Green's function. We
have suppressed the $a$ index in places.

The ground state wavefunction for the Hamiltonian
(\ref{eq:hamiltonian}) will now be shown to be
\be\label{eq:groundstate}
\psi_0 = e^{- \sum_i K_i} \,,
\ee
where $K_i = K(z_i, \zb_i)$. Acting on this state with the
Hamiltonian (\ref{eq:hamiltonian}) gives
\be\label{eq:formula}
H \psi_0 = \sum_j \left( K_j + 3 - \frac{g^{a \bar b}_j}{2}
\left[ (\nabla_{z_j^a} K_j) \nabla_{\zb_j^{\bar b}} +
(\nabla_{\zb_j^{\bar b}} K_j ) \nabla_{z_j^a}
\right] (K_j + 2 \sum_{k \neq j} s_{kj})
\right) \psi_0
\,.
\ee
Now, for an arbitrary Calabi-Yau cone with metric
(\ref{eq:metric}) we have that
\be
K = \frac{r^2}{2} \,.
\ee
This can be derived from a short argument starting with the
observation that the K\"ahler form is homogeneous with degree two
in $r$, see e.g. \cite{Martelli:2005tp}. It follows that the
vector appearing in (\ref{eq:formula}) is the Euler vector of the
cone
\be\label{eq:euler}
g^{a \bar b}_j
\left[ (\nabla_{z_j^a} K_j) \nabla_{\zb_j^{\bar b}} +
(\nabla_{\zb_j^{\bar b}} K_j ) \nabla_{z_j^a}
\right] = r \frac{\pa}{\pa r} \,.
\ee
The scaling (\ref{eq:scaling}) then implies that
\be
\sum_i r_i \frac{\pa}{\pa r_i} \sum_{j \neq i} s_{ij} =
- \frac{N(N-1)}{2}\frac{\pi^3}{\text{Vol}(X_5)} \,.
\ee
Putting the above statements together we obtain
\be
H \psi_0 = \left(3 N +
\frac{N(N-1)}{2}\frac{\pi^3}{\text{Vol}(X_5)}
\right) \psi_0 \equiv E_0 \psi_0\,.
\ee
Thus $\psi_0$ is an eigenstate as claimed. The lack of dependence
on the angular coordinates $\theta$ suggests that it is the ground
state. The two key ingredients here were the relation between the
K\"ahler potential and the Euler vector (\ref{eq:euler}), and the
scaling behaviour of the Green's function (\ref{eq:scaling}). Any
scaling function would have given the same results.

\section{The emergent geometry}

In the large $N$ limit, the ground state wavefunction
(\ref{eq:groundstate}) describes an emergent semiclassical
geometry \cite{Berenstein:2005aa}. This occurs because a specific
configuration of eigenvalues dominates the matrix integral.

The probability of the eigenvalues being in some particular
distribution is given by the square of the wavefunction multiplied
by the measure factor (\ref{eq:measure}) needed to make the
Hamiltonian (\ref{eq:hamiltonian}) self-adjoint. That is
\be
\mu^2 |\psi_0|^2 = e^{- \sum_i r_i^2 - \sum_{j \neq i} s_{ij}} \equiv e^{-S} \,.
\ee
In the large $N$ limit, we expect a particular configuration to
dominate. This will be given by minimizing the effective action
\be
S = \int d^6x \rho(x) r_x^2 + \int d^6x d^6y \rho(x) \rho(y)
s(x,y)
\,,
\ee
where we have introduced the large $N$ eigenvalue density,
$\rho(x)$, which satisfies
\be\label{eq:constraint}
\int d^6x \rho(x) = N \,.
\ee
The notation we are using here is that $x$ runs over the six
coordinates on the cone, which we denote $r_x$ and $\theta_x$.

The saddle point equations are
\bea
r_x = - \int dr_y d\q_y r_y^5 \sqrt{g_5(\q_y)} \rho(r_y,\q_y)
\frac{\pa s(r_x,r_y,\q_x,\q_y)}{\pa r_x}
\,, \\
\quad 0 = \int dr_y d\q_y r_y^5 \sqrt{g_5(\q_y)} \rho(r_y,\q_y)  \frac{\pa s(r_x,r_y,\q_x,\q_y)}{\pa
\q_x}\,,
\eea
where we have explicitly separated the dependence on the $r$ and
$\q$ coordinates.

With the Green's function discussed above, one can prove that the
density of eigenvalues is not smooth. This is a generalization of
an argument found in \cite{Berenstein:2005aa}. The basic idea is
that we can also write the saddle point equations as
\be
K(x) + \int d^6y \rho(y) s(y,x) = C \,, \label{eq:lagrange}
\ee
where $C$ is a Lagrange multiplier enforcing the constraint in
equation (\ref{eq:constraint}) and $K$ is the K\"ahler potential.
From here, if $\rho$ is smooth, the operation of differentiating
with respect to $x$ commutes with the integral. We can act with
the Laplacian associated to the metric (\ref{eq:metric}) three
times on both sides of the equation. On the left hand side we find
that $\nabla^2 K$ is constant, and further applications of
$\nabla^2$ give zero. Inside the integral, we would act three
times with $\nabla^2$ on the Green's function, and then we would
use the defining equation of the Green's function itself to find
that $\rho=0$. This is incompatible with the constraint, so the
assumption that $\rho$ has smooth support is wrong. The simplest
solution that one could imagine with singular support will have
some $\delta$ function distribution in it.

Using the formulae in Appendix \ref{sec:log} for $s$, equations
(\ref{eq:epsilon}) - (\ref{eq:result}), and the fact that $\int
d\q \sqrt{g_5} \Theta_\n(\q) = 0$ for $\nu > 0$, i.e. that the
higher harmonics on $X_5$ integrate to zero, it is straightforward
to see that if $\rho(r,\q)$ has no $\q$ dependence, then all the
$\theta_x$ dependence drops out of the equations of motion once
the $\theta_y$ integrals are done. Solving the saddle point
equations reduces to a purely radial problem. Moreover, since we
know that the density of eigenvalues has singular support, we can
make a simple guess to solve the problem.

The eigenvalues are found to fill out an $X_5$:
\be\label{eq:solution}
\rho(x) = \frac{N \d(r_x-r)}{r_x^5 \text{Vol}(X_5)} \,,
\ee
at the constant radius
\be\label{eq:radius}
r = \sqrt{\frac{N}{2}} \sqrt{\frac{\pi^3}{\text{Vol}(X_5)}} \,.
\ee
This expression reduces to the previously known $S^5$ of radius
$r=\sqrt{N/2}$ when the cone is over a sphere, as $\text{Vol}(S^5)
= \pi^3$. It is a solution to the equations of motion for all base
manifolds $X_5$, it does not depend on the manifold being
homogeneous. Thus we see that part of the $AdS_5 \times X_5$
geometry has emerged from the eigenvalue quantum mechanics. We
obtain $X_5$ together with its Sasaki-Einstein metric, because of
the requirement that the metric on the conical target space of the
eigenvalues is Calabi-Yau \cite{Berenstein:2007wi}.

This $X_5$ eigenvalue distribution is to be understood as the
large $N$ ground state of the theory, where quantum mechanical
measure effects have repelled the eigenvalues away from their
classical origin at $r = 0$. It is a self consistent starting
point for studying the low energy dynamics. All off diagonal
fluctuations are massive
\cite{Berenstein:2005aa, Berenstein:2005jq} as are all the higher
harmonics on $S^3$ (this second point follows from the analysis in
\cite{Gursoy:2007np, Aharony:2007rj}). In the remainder of this
paper we study three particular excitations about this ground
state. We will see that they reproduce quantitative features of
strings and D branes in the dual spacetime.

\section{The spectrum of scalar Kaluza-Klein harmonics}

\subsection{The spectrum for ${\cal N}=4$}

In the ${\cal N}=4$ SYM theory, the spectrum of gravity multiplets
can be deduced from the half BPS states. The half BPS primary
fields corresponding to single graviton states are given by
single-trace operators of the form $\Tr z^n$, with $z = x^1 + i
x^2$. These are holomorphic highest weight states of $SO(6)$, for
a symmetric traceless tensor representation of $SO(6)$.

In the commuting matrix model of strong coupling, as we reviewed
in section 2 above, the wave functions of these states are
conjectured to be
\begin{equation}
\psi = \psi_0 \Tr z^n \,,
\end{equation}
where $\psi_0$ is the ground state wave function of the matrix
model. In the ${\cal N}=4$ matrix model, on $\R^6$, the $SO(6)$
symmetry is manifestly part of the dynamics, and $\psi_0$ is an
$SO(6)$ singlet. It is natural to expect that the wave functions
of other states that are not half BPS with respect to the same
half of the supersymmetries as $z^n$ are given by
\begin{equation}
\psi = c_{i_1 \dots i_n}\Tr(x^{i_1}\dots x^{i_n})\psi_0 \,,
\end{equation}
where $c_{i_1 \dots i_n}$ is symmetric and traceless.

Since the matrices commute, the trace is just a sum over
eigenvalues, and we find ourselves with a one-particle wave
function problem. The resulting symmetric traceless polynomials of
six variables are characterized by the property that
\begin{equation}
\nabla^2(c_{i_1 \dots i_n} x^{i_1}\dots x^{i_n}) \sim c_{i_1 i_1 i_3 \dots i_n} x^{i_3}\dots
x^{i_n}=0\,,
\label{eq:harm}
\end{equation}
this is, they are harmonic functions on $\R^6$.

These states have energy $n$, and thus the dual operators have
dimension $n$. We can recover this result by considering the
one-particle wave function problem for a six-dimensional harmonic
oscillator
\begin{equation}\label{eq:single}
H= -\frac 12 \nabla^2+ \frac 12 \vec x^2 \,.
\end{equation}
This Hamiltonian differs from the full multi matrix model
Hamiltonian (\ref{eq:hamiltonian}) for the ${\cal N}=4$ problem by
the absence of the measure, which mixes the eigenvalues. We show
in Appendix B, for the general conifold, that the measure may be
neglected for these states to leading order at large $N$. Thus in
this limit it is sufficient to investigate the spectrum of
(\ref{eq:single}). The absence of mixing between eigenvalues
allows us to focus on a single eigenvalue, hence we have dropped
the $i$ index in (\ref{eq:single}).

We take our wave function to be
\begin{equation}
\psi= e^{- \vec x^2/2} c_{i_1 \dots i_n} x^{i_1}\dots
x^{i_n}\label{eq:psisym}\,.
\end{equation}
When we calculate $\nabla^2\psi$, there are three types of terms
that appear. The terms with two derivatives acting on the
exponential are cancelled by the term $\frac 12 x^2$ in the
Hamiltonian. The terms with two derivatives acting on the
polynomial vanish because of equation (\ref{eq:harm}). We are left
with terms with one derivative acting on the exponential and one
derivative acting on the polynomial. If we write the Laplacian in
spherical coordinates, we find that these terms give
\begin{equation}
 c_{i_1 \dots i_n} e^{- r^2/2} \left(\frac{1}{2 r^5} \frac{d}{dr} \left( r^6
x^{i_1}\dots x^{i_n}\right) +
\frac{r}{2} \frac{d}{dr} \left(x^{i_1}\dots x^{i_n}\right) \right) \,.
\end{equation}
Now, $x^i = r f^i(\theta)$, for some $f^i(\theta)$, so we need to
evaluate
\begin{equation}
\frac{1}{2 r^5} \frac{d}{dr} r^{n+6} + \frac{r}{2} \frac{d}{dr} r^n = (n+3)
r^n\,.
\end{equation}
Via this exercise, we find that the wavefunction written down in
equation (\ref{eq:psisym}) is an eigenfunction of the one particle
Hamiltonian (\ref{eq:single}), and that its energy is $n$ units
greater than the energy of the vacuum state. The same value of $n$
is the dimension of the corresponding operator in the conformal
field theory. This calculation provides a link between the energy
of a state in the harmonic oscillator problem, and the dimension
of the corresponding state in supergravity. We should also notice
that what matters for this computation is that the polynomial we
considered was a homogeneous function (it is a scaling function
under the vector $r\partial_r$), and that the energy obtained is
exactly this scaling dimension.

In the case of ${\cal N}=4$ SYM, all of these symmetric traceless
functions are obtained by acting with rotations on $z^n$, and
therefore they are in some sense locally holomorphic with respect
to a suitable choice of complex coordinates. This is characterized
exactly by having a harmonic function. We will now extend this
calculation on the moduli space of a `single brane' in ${\cal N}=4
$ SYM to the case of a `single brane' in the case of a conformal
field theory associated to a general conifold. As we noted, the
term mixing the eigenvalues in the Hamiltonian, the measure, will
again not be important to leading order at large $N$ for this
problem.

\subsection{The spectrum for general conifolds}

As we have discussed, for the general conifold the eigenvalue
dynamics is locally given by ${\cal N}=4 $ SYM. The wave function
is a global object, but the property of being a harmonic function
is something that one can check locally, as it is governed by a
second order differential equation. It seems natural to take the
same ansatz for this more involved case as we did for ${\cal N}=4
$ SYM. We consider wave functions of the form
\begin{equation}\label{eq:wavefn}
\psi = \psi_0 \Tr h(x) \,,
\end{equation}
where $h(x)$ is a harmonic function on the Calabi-Yau cone over
$X_5$ and $\psi_0 = e^{-r^2/2}$, as we found above.

The one particle problem (i.e. without the measure, see Appendix
B) now corresponds to the Hamiltonian
\begin{equation}
H= -\frac 1{2  r^5} \frac{d}{dr} r^5 \frac{d}{dr} -\frac{1}{2 r^2}
\nabla_5^2 +\frac{r^2}{2} \,.
\end{equation}
One can separate variables in $\theta$ (the coordinates on $X_5$)
and $r$, and hence consider harmonic functions of the form $h(x) =
h(r) \Theta(\theta)$, where $\Theta$ is an eigenfunction of the
Laplacian on the Sasaki-Einstein space. That is
\begin{equation}
- \nabla_5^2 \Theta(\theta) = \nu^2 \Theta(\theta) \,.
\end{equation}
Harmonicity now requires solving the following differential
equation for $h(r)$
\begin{equation}
\left(-\frac{1}{r^5} \frac{d}{dr} r^5 \frac{d}{dr} +\frac {\nu^2}{r^2}\right)
h(r)=0 \,.
\end{equation}
This is solved by $h(r) = r^\lambda$, where $\lambda$ satisfies
\begin{equation}
\lambda(\lambda+4)- \nu^2=0\label{eq:harm2} \,,
\end{equation}
or equivalently
\begin{equation}
\lambda= -2+\sqrt{4+\nu^2} \,,
\end{equation}
where we chose the root that makes the wavefunction nonsingular at
the origin. The same manipulations that told us in the case of
${\cal N}=4$ SYM that the energy of the harmonic function of
weight $n$ multiplying the ground state wave function has energy
$n$, now show us that the energy of the single-particle wave
function (\ref{eq:wavefn}) is given by $\lambda$.

The equation (\ref{eq:harm2}) is familiar from supergravity in
$AdS_5$ \cite{Gubser, Nieuw} (see also \cite{Cer}), where one
associates a scaling dimension $\lambda$ to a scalar particle in
five dimensions that originates from perturbations mixing the
graviton and the self-dual five-form field strength. We see that
the scaling dimensions of the operators are controlled by harmonic
analysis on the Sasaki-Einstein space, recovering exactly the
spectrum of some of the scalar fluctuations in the dual gravity
theory. In particular, for all holomorphic wave functions we
recover the exact scaling dimension predicted by the chiral ring.
Most of these harmonic functions are not holomorphic, however, so
we are recovering universally the spectrum of a large family of
non-BPS Kaluza-Klein harmonics of the dual supergravity theory.

In Appendix C we discuss the possibility of building coherent
states using these single trace states. These appear to be dual to
classical geometries, as one would expect for coherent states of
gravitons.

\section{Spectrum of off-diagonal fluctuations}

The off diagonal modes connect pairs of eigenvalues. For small
separations, $\Delta z_{ij} = z_j - z_i$, the energies of these
modes are given by their mass term plus the distance between the
two eigenvalues, see \cite{Berenstein:2005jq, Berenstein:2007wi}
and section 2 above,
\be
E^2_{ij} = 1 + \frac{g^2_\text{eff.}}{2 \pi^2} g_{i \, a \bar b} \Delta z^a_{ij} \Delta \zb^{\bar b}_{ij}  \,.
\ee
Recall that $g_{\text{eff.}}$ is the effective ${\mathcal{N}}=4$
coupling which controls the masses of off diagonal modes
connecting nearby eigenvalues. The $z_i$ are all at constant
radius r given by (\ref{eq:radius}). Thus we have
\be\label{eq:Eij}
E^2_{ij} = 1 + \frac{\lambda_\text{eff.}}{4 \pi^2}
\frac{\pi^3}{\text{Vol}(X_5)} | \Delta \theta_{ij} |^2_{g_{5,i}}
\,,
\ee
where $\lambda_\text{eff.} = g^2_\text{eff.} N$, $\Delta
\theta_{ij}$ is the separation in $X_5$, and $g_5$ is the metric on $X_5$.

We would like to write down an operator that describes these off
diagonal fluctuations. The operators that do the trick
\cite{Berenstein:2005aa, Berenstein:2005jq, Berenstein:2007zf} are
strong coupling realisations of the BMN \cite{Berenstein:2002jq}
operators
\be\label{eq:operator}
{\mathcal{O}}_{k,J} = \sum_{n=1}^J \Tr \left[ z^n \b^\dagger
z^{J-n} \tilde \beta^\dagger \right] e^{2 \pi i n k/J} \,.
\ee
In this expression $k$ is an integer, $J$ is the R charge of the
operator, $\b^\dagger$ and $\tilde \b^\dagger$ are creation
operators for off diagonal modes, and $z$ is a complex coordinate
on the conical moduli space with a fixed scaling dimension $c$.

The wavefunction corresponding to this operator is
\be\label{eq:psikJ}
\psi_{k,J} = {\mathcal{O}}_{k,J} \psi_0 \,.
\ee
In principle, the inclusion of the operator ${\mathcal{O}}_{k,J}$
will backreact on the dominant eigenvalue distribution, in a way
similar to that described below in probing the radial direction.
However, here we wish to take $J$ large, but not of order $N$. In
this case the effect of the insertion of ${\mathcal{O}}_{k,J}$ in
(\ref{eq:psikJ}) on the eigenvalue distribution is subleading in
$1/N$. Thus we can take the eigenvalues $z_i$ to lie on the ground
state solution (\ref{eq:solution}).

Invariance under the unbroken $U(1)^N$ symmetry requires that
$\b^\dagger$ and $\tilde \b^\dagger$ carry opposite charges. Thus
if we take $\b^\dagger$ to connect the $i$th and $j$th
eigenvalues, then $\tilde \b^\dagger$ must connect the $j$th to
the $i$th. This is implemented automatically by the trace in
(\ref{eq:operator}). The operator (\ref{eq:operator}) may be
written
\be\label{eq:nsum}
{\mathcal{O}}_{k,J} = \sum_{i,j} \sum_{n=1}^J z^n_i z^{J-n}_j
\b^\dagger_{ij}
\tilde \beta^\dagger_{ji} e^{2 \pi i n k/J} \,.
\ee
At large $J$, there is a dominant contribution to this sum
\cite{Berenstein:2005jq, Berenstein:2007zf}. Firstly, the
dominantly contributing eigenvalues maximize $|z|$. This does not
fix the location along the angle $\psi$ dual to the R charge, as
$z$ is a chiral operator and hence $|z|$ is invariant under R
charge rotations. More specifically, on the locus where $|z|$ is
maximized we may write
\be\label{eq:rpsi}
z_i \propto r^c e^{i c \psi_i} \,.
\ee
The exponent follows from two observations. Firstly, because $z$
has conformal dimension $c$, we have $r \pa_r z = c z$. Secondly,
see for instance \cite{Gibbons:2002th, Martelli:2006yb}, $\pa_\psi
= {\mathcal{J}} (r \pa_r)$, where $\mathcal{J}$ is the complex
structure on the Calabi-Yau. Therefore $\pa_\psi z = i c z$, as
implied by (\ref{eq:rpsi}). Now doing a saddle point approximation
to the sum over $n$ in (\ref{eq:nsum}) we find
\be\label{eq:distance}
\psi_i - \psi_j = - \frac{2 \pi k}{c J}\,.
\ee
This is a crucial relation which says that for given $k$ and $J$,
the dominant contribution to the operator ${\mathcal{O}}_{k,J}$
comes from two off diagonal modes connecting a pair of eigenvalues
separated according to (\ref{eq:distance}). It follows from our
previous expression (\ref{eq:Eij}) for the off diagonal energies
that
\be\label{eq:energies}
E_{{\mathcal{O}}_{k,J}} - c J = 2 E_{ij} = 2 \sqrt{1 +
\frac{\lambda_\text{eff.} \pi^3}{\text{Vol}(X_5)} \frac{k^2}{c^2 J^2}} \,,
\ee
where we included the contribution to the energy from $z^J$ in
(\ref{eq:operator}). Conveniently, we did not need to find the point
on the remaining directions in $X_5$ at which $|z|$ is maximized, as
$g_5(\pa_\psi,\pa_\psi)=1$ is in fact constant over the
Sasaki-Einstein space, see for instance
\cite{Gibbons:2002th, Martelli:2006yb}. We will now see that this
result is precisely the spectrum of excitations about a rapidly
rotating BMN string in the dual spacetime.

\section{Comparison with the plane wave limit}

The spacetime dual to the superconformal field theory is $AdS_5
\times X_5$. The metric may be written as
\be
ds^2 = L^2 ds^2_{AdS_5} + L^2 \left[(d\psi +
\sigma)^2 + ds^2_{KE} \right] \,.
\ee
Here $ds^2_{KE}$ is a four dimensional K\"ahler-Einstein metric
and $d \sigma$ is proportional to the K\"ahler two form
corresponding to this metric. We restrict ourselves here to
(quasi)regular Sasaki-Einstein manifolds, in which the fibre
coordinate $\psi$ has a finite periodicity.

States with large angular momentum about the $\psi$ direction,
corresponding to large $R$ charge, are captured by the Penrose
limit of this background \cite{Berenstein:2002jq}. This limit was
computed in \cite{Itzhaki:2002kh} -- Penrose limits of the special
case of $X_5 = T^{1,1}$ were also computed in
\cite{Gomis:2002km,Pando Zayas:2002rx} -- to give
\be\label{eq:pp}
ds^2 = - 4 dx^+ dx^- - |x|^2_8 (dx^+)^2 + dx^2_8 \,,
\ee
where
\be\label{eq:coords}
x^+ = \frac{1}{2} ( t + \psi) \,, \qquad x^- =
\frac{L^2}{2} ( t - \psi) \,.
\ee
Note that (\ref{eq:pp}) is just the maximally supersymmetric plane
wave background \cite{Berenstein:2002jq, Blau:2002dy}.

The conformal dimension and R charge are given by
\be
\Delta = i \pa_t \,, \qquad c J = - i \pa_\psi \,.
\ee
Where for ease of comparison with the previous subsection, we denote the total
R charge by $c J$. Therefore from (\ref{eq:coords})
\bea\label{eq:momenta}
2 p^- & = & i \pa_{x^+} = i \left(\pa_t +
\pa_\psi \right) = \Delta - c J \,, \\
2 p^+ & = & i \pa_{x^-} = \frac{i}{L^2} \left(\pa_t -
\pa_\psi \right) = \frac{1}{L^2} \left( \Delta + c J \right) \,.
\eea
Quantising the string excitations \cite{Berenstein:2002jq,
Metsaev:2001bj} in the plane wave background (\ref{eq:pp}) gives
the spectrum of excitations
\be
2 \delta p^- = \sqrt{1 + \frac{k^2}{\a'^2 (p^+)^2}} \,.
\ee
Using the expressions for the momenta (\ref{eq:momenta}) and
working to leading order at large $J$, but with $L^2/\a' J$ fixed,
one obtains
\be\label{eq:stringlevels}
\Delta - c J =
\sqrt{1+ \frac{L^4 k^2}{\a'^2 c^2 J^2}} \,.
\ee

The supergravity background has a Ramond-Ramond five form
\be
F^{(5)} = \frac{N \sqrt{\pi}}{2 \text{Vol}(X_5)}
\left(\text{vol}_{AdS_5} + \text{vol}_{X_5} \right) \,.
\ee
The solution to the supergravity equations specifies a relation
between the units of flux, $N$, and the AdS radius, $L$, in string
units
\be
\frac{L^4}{\a'^2} = \frac{4 \pi g_s N \pi^3}{\text{Vol}(X_5)} \,,
\ee
To further relate this expression to our previous results, note
that the local effective ${\mathcal{N}}=4$ coupling,
$g_{\text{eff.}}$, must be related to the expectation value of the
dilaton in the usual way
\be
g^2_\text{eff.} = 4 \pi g_s \,.
\ee
This follows, for instance, by noting that these quantities
transform in the correct way under S duality. Hence we obtain from
(\ref{eq:stringlevels})
\be\label{eq:finalresult}
\Delta - c J =
\sqrt{1+\frac{\lambda_\text{eff.} \pi^3}{\text{Vol}(X_5)} \frac{k^2}{c^2 J^2}} \,.
\ee
Recalling that the eigenvalue Hamiltonian is in fact the conformal
dimension, $H = \Delta$, we have
precisely reproduced the matrix quantum mechanics result
(\ref{eq:energies}). We need to multiply (\ref{eq:finalresult}) by
two because we are considering two excitations. Note that we
nontrivially match the appearance of the volume factor
$\text{Vol}(X_5)$.

\section{Exciting an eigenvalue: the radial direction}

In the previous two sections we have shown how off diagonal modes
connecting ground state eigenvalues are dual to string excitations
about the $AdS_5 \times X_5$ background in the BMN limit. In this
section we return to purely eigenvalue excitations, no
off-diagonal modes, but with a larger R charge, $J
\sim N$. We will see how these excitations move a single
eigenvalue into the radial direction, and are dual to AdS giant
gravitons. To familiarise ourselves with the procedure, we will
consider the ${\mathcal{N}}=4$ case first.

\subsection{Probing the radial direction in ${\mathcal{N}}=4$}

In the ${\mathcal{N}}=4$ case, the cone is over $S^5$, i.e. the
total space is just $\R^6$. We will use cartesian coodinates $\vec
x$ to denote the matrices, rather than the `polar' coordinates
$r,\q$.

Consider the wavefunction
\be
\psi = \psi_0 \Tr z^J \,,
\ee
where $z = x^1 + i x^2$ and $\psi_0$ is the ground state
wavefunction. In Appendix B we show that in the large $N$ limit,
this is an eigenfunction of the Hamiltonian (\ref{eq:hamiltonian})
with eigenvalue $E=E_0 + J$. The probability density is
\be\label{eq:density}
\mu^2 |\psi|^2 = e^{- \sum_i \vec x_i^2 + \frac{1}{2} \sum_{i \neq j} \log
|\vec x_i - \vec x_j|^2 + \log \sum_{i,j} \left(x^1_i + i x^2_i
\right)^J \left( x^1_j - i x^2_j \right)^J} \,,
\ee
where we used the explicit Green's function on $\R^6$ to write
$\mu^2 = \prod_{i < j} |\vec x_i - \vec x_j|^2$.

We will make the assumption, to be verified a posteriori, that
$|x_N| > |x_i|$ for all $i \neq N$ and that $J \gg 1$. We may thus
approximate the last term
\be
 \log \sum_{i,j} \left(x^1_i + i x^2_i
\right)^J \left( x^1_j - i x^2_j \right)^J \to
J \log \left[ (x_N^1)^{2} + (x_N^2)^{2} \right] \,.
\ee
This is the assumption that one eigenvalue will be moved away from
the others.

The large $N$ semiclassical support of the wavefunction is found
by extremising the exponent in (\ref{eq:density}). The equations
of motion are
\be\label{eq:eom}
x^A_i = \sum_{i \neq j} \frac{x^A_i - x^A_j}{|\vec x_i -
\vec x_j|^2} + \frac{J \d_{iN} \left[\d^{1A} x_N^1 + \d^{2 A}
x_N^2 \right]}{(x_N^1)^{2} + (x_N^2)^{2}} \,.
\ee
We look for a solution to these equations which is given by the
ground state before the insertion, an $S^5$ of radius squared $r^2
= N/2$ and density $N/\pi^3 r^5$, together with a single
eigenvalue $\vec x_N$ separated from the sphere. There will be an
$S^1$ worth of such solutions, where the $S^1$ lies in the
$x^1-x^2$ plane. Without loss of generality, we can take the
eigenvalue to move off in the $x^1$ direction
\be
x_N^A = x_N \d^{1A} \,.
\ee
For $i \neq N$, the equation of motion (\ref{eq:eom}) is satisfied
to leading order in $N$, because the equation of motion is just
that corresponding to the ground state wavefunction which is
solved by the $S^5$. The effect of the extra eigenvalue is
subleading. The equation for $i=N$, however, gives a nontrivial
equation for $x_N$. Using the integral
\be
\frac{8 N}{3 \pi} \int_0^\pi d\q \sin^4 \q
\frac{x_N - r \cos \q}{x_N^2 + r^2 - 2 x_N r \cos\q} = \frac{N (6 x_N^4 - 4 x_N^2 r^2 + r^4)}{6 x_N^5} \,,
\ee
the equation of motion becomes, using $r^2 = N/2$,
\be\label{eq:xn}
x_N^6 - (J+N) x_N^4 + \frac{N^2}{3} x_N^2 - \frac{N^3}{24} = 0 \,.
\ee
If we set
\be
x_N^2 = d^2 N \,, \qquad J = j N \,,
\ee
then the solution to the (cubic) equation (\ref{eq:xn}) is
\be\label{eq:d2}
d^2 = \frac{1+ j}{3} + \frac{2 j (j+2)}{3 p(j)^{1/3}} +
\frac{p(j)^{1/3}}{6} \,,
\ee
where
\be
p(j) = \frac{1}{2} \left[ 1 + 24 j + 48 j^2 + 16 j^3 + \sqrt{1 +
48 j + 672 j^2 + 288 j^3} \right] \,.
\ee
This gives us the distance of the $x_N$ eigenvalue from the origin
as a function of the angular momentum $J$. We see that $J \sim N$
is indeed large as required. Figure 1 illustrates the
configuration we have just obtained.

\begin{figure}
\centerline{ \epsfig{figure=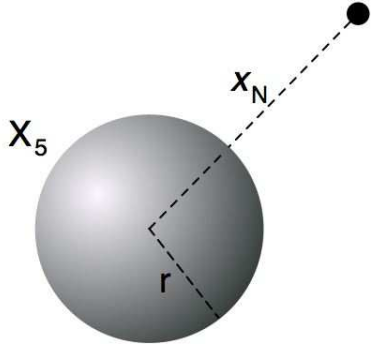, width=5cm} }
\caption{A single eigenvalue is removed from the $X_5$ to a distance $x_N \sim \sqrt{J}$.}
\end{figure}

Taking the further limit that the eigenvalue is far away from the
sphere in $\sqrt{N}$ units, i.e. $j \gg 1$, gives the result
\be
x_N = \sqrt{J} + \cdots \qquad (J/N \gg 1) \,.
\ee

The association of an object with large R charge to a radial
motion is strongly reminiscent of AdS giant gravitons. This will
shortly lead us to identify the radial direction of the Calabi-Yau
cone outside of the $X_5$ occupied by the ground state with the
radial direction of global $AdS_5$.

\subsection{Probing the radial direction for general conifolds}

The argument goes through essentially unchanged for the case of a
general cone over $X_5$. We make the assumption that one
eigenvalue will have a larger modulus than the others $|z_N| >
|z_i|$, for all $i \neq N$. Thus in the limit $J \gg 1$ we may
write the probability density as
\be\label{eq:generalaction}
\mu^2 |\psi|^2 = \mu^2 |\psi_0 \Tr z^J|^2 = e^{- \sum_i r_i^2 - \sum_{i \neq j} s_{ij} + J \log
|z_N|^2}\,.
\ee
As we note in Appendix B, the holomorphic coordinate $z$ must be a
power of $r$ multiplied by a harmonic function on $X_5$
\be
z = r^c F_c(\theta) \,.
\ee
The large $N$ semiclassical equations of motion following from
(\ref{eq:generalaction}) are therefore
\be
r_i + \sum_{j \neq i} \frac{\pa s_{ij}}{\pa r_i} = \frac{c J
\d_{iN}}{r_N} \,, \qquad \sum_{j \neq i} \frac{\pa s_{ij}}{\pa \q_i} = \frac{J
\d_{iN} \pa_{\q_N} |F_c(\q_N)|}{|F_c(\q_N)|} \,.
\ee

In the large $N$ limit, as for the case of $S^5$ above, the
equations of motion for the eigenvalues $i \neq N$ are unaffected
by the insertion of $\Tr z^J$, as the motion of the single eigenvalue
$z_N$ away from the ground state configuration is a subleading
effect. The equations of motion for $r_N$ and $\q_N$ however are
nontrivial. Recall the observation we made in section 4: that the
independence of the ground state eigenvalue density on $\theta$
implies that any integral of the form $\int d\q_x
\rho(\theta_x) s(\theta_x,\q_y)$
kills the $\theta_y$ dependence. This fact, together with the
expression for $s$ in equation (\ref{eq:epsilon}) and the integral
\be
\int_0^\infty \frac{d \lambda}{\lambda^3} \frac{J_2(\sqrt{\lambda}
r)}{r^2}\frac{\pa}{\pa r_N} \frac{J_2(\sqrt{\lambda} r_N)}{r_N^2}
= \frac{- 4 r_N^2 r^2 + 6 r_N^4 + r^4}{192 r_N^5} \,,
\ee
leads to the following equations
\bea
& r_N^6 - \left(c J + \frac{N \pi^3}{\text{Vol}(X_5)} \right)
r_N^4 +
\frac{1}{3} \left(\frac{N \pi^3}{\text{Vol}(X_5)}\right)^2 r_N^2 -
\frac{1}{24} \left(\frac{N \pi^3}{\text{Vol}(X_5)}\right)^3 = 0  \,,  \label{eq:radeq} \\
& \pa_{\q_N} |F_c(\q_N)| = 0 \,, \label{eq:thetaeq}
\eea
where we also used the radius of the ground state $X_5$ in (\ref{eq:radius}).

The immediate observation we can make from these equations is that
the radial and angular parts have completely decoupled. We can
interpret this as the fact that the radial direction in the bulk
geometries emerges universally. It does not depend on where the
eigenvalue is sitting in $X_5$. This reflects the direct product
structure of the dual geometry: $AdS_5 \times X_5$.

The equation (\ref{eq:thetaeq}) for $\q$ says that $|z_N|$ is
maximized given its fixed radius $r_N$. This, together with the
fact that we will find $r_N > r$, guarantees that our assumption
that $|z_N| > |z_i|$ for $i \neq N$ is consistent. As in the case
for $S^5$, there will not be a unique solution to
(\ref{eq:thetaeq}). Rather there will be an $S^1$ worth of
solutions, corresponding to the R symmetry circle.

If we make the definitions
\be
r_N^2 = d^2 \frac{N \pi^3}{\text{Vol}(X_5)} \,, \qquad c J = j
\frac{N \pi^3}{\text{Vol}(X_5)} \,,
\ee
then we find that the radial equation (\ref{eq:radeq}) is exactly
the same as the one we found in the case of $S^5$, with solution
(\ref{eq:d2}). Thus (\ref{eq:d2}) describes how the eigenvalue
$z_N$ moves out in the radial direction as a function of the R
charge $c J$. From the equation (\ref{eq:radeq}) we see that the
general relation between $r_N$ and $cJ$ depends on the volume of
$X_5$. However, in the limit $j
\gg 1$ we again find
\be\label{eq:rn}
r_N = \sqrt{c J} + \cdots \quad (J/N \gg 1) \,.
\ee

\subsection{Comparison with AdS giant gravitons}

In the ${\mathcal{N}}=4$ case, at weak coupling, the wavefunction
dual to an AdS giant graviton with angular momentum $J$ along the
equator of the $S^5$ is \cite{Corley:2001zk}
\be
\psi = \psi_0 \chi_{S_J}(z) \,,
\ee
where $\chi_{S_J}$ is the Schur polynomial corresponding to the
totally symmetric representation of rank $J$. In terms of the
eigenvalues of $z$
\be\label{eq:schur}
\chi_{S_J}(z) = \sum_{1 \leq i_1 \leq \cdots \leq i_J \leq N}
z_{i_1} \cdots z_{i_J} \,.
\ee
We would like to approximate $\chi_{S_J}(z)$ with $\Tr z^J$, so
that we can use the results of the previous section to evaluate
the semiclassical wavefunction. This will be valid provided that
the largest eigenvalue $|x_N/x_p| \gg 1$ for all $p \neq N$, which
requires $j \gg 1$. In this limit $d^2 = j$ in (\ref{eq:d2}).
Furthermore, it is unclear that the Schur polynomials
(\ref{eq:schur}) will be orthogonal at strong coupling. On the
other hand, we have shown in Appendix B that the states $\psi_0
\Tr z^J$ are eigenstates to leading order at large $N$, with
different eigenvalues, and therefore are orthogonal.

In the bulk, the AdS giant gravitons are $D3$ branes in which an
$S^3$ expands to a finite radius in $AdS_5$, due to their angular
momentum about the R symmetry direction. With angular momentum $J$
the radius is
\cite{Grisaru:2000zn, Hashimoto:2000zp} given by
$r^2_{\text{giant}} = J/N$. Here we are using global coordinates
in $AdS_5 \times X_5$
\be\label{eq:ads}
\frac{1}{L^2} ds^2 = - (1+r^2) dt^2 + \frac{dr^2}{1+r^2} +
r^2 d\Omega^2_{S^3} + d \Omega^2_{X_5} \,.
\ee

Comparing this bulk result with our matrix model result
(\ref{eq:rn}), and absorbing the factor of $c$ into the definition
of $J$, now the total R charge, we obtain
\be
\frac{r_N}{\sqrt{N}} = r_{\text{giant}} \qquad \text{for} \qquad r_{\text{giant}}
\gg 1 \,.
\ee
Thus the radial direction in the space of eigenvalues, in units of
$\sqrt{N}$, is exactly equal to the radial direction of $AdS$ in
the large radius limit. This is a nice result, but it is also not
clear why the particular coordinate $r$ that we chose in
(\ref{eq:ads}) should have been singled out in this way.

We can unambiguously draw the following conclusions from this
section, however: The radial direction of the Calabi-Yau cone
outside of the $X_5$ occupied by the eigenvalues is to be
topologically identified with the radial direction in global
$AdS_5$. This direction emerges independently of and orthogonally
to the manifold $X_5$. The radial coordinate may be probed using
operators with large R charge and identifying the dual string
theory states. It is of great interest to obtain more
information using this approach, such as the warping of spacetime
and the redshift of $AdS$ as a function of radius.

\section{Discussion and future directions}

One main objective of this paper has been to show that a framework
now exists for performing precise computations in many strongly coupled
${\mathcal N} = 1$ conformal field theories. This is of interest
because these theories are dual to compactifications of type IIB
string theory on Sasaki-Einstein spaces. We have seen how the
Sasaki-Einstein manifold emerges as the semiclassical limit of the
ground state of a commuting matrix model. We have then found that
the spectrum of certain non-BPS supergravity and stringy
excitations may be reproduced exactly as excited states in the
matrix quantum mechanics.

The basic setup has exploited a connection that all these field
theories have an effective (local) ${\cal N}=4$ SYM description on
moduli space, and that one should copy strategies that worked in
${\cal N} =4$ SYM by analogy and a use of local concepts on moduli
space. In particular, the formalism used in this paper required
the introduction of a measure that was determined by solving a
differential equation on the moduli space. If one can find a
closed form expression for the corresponding measure in various
cases (let us say the conifold), it would be possible to test this
proposal further.

A very important result that follows from our proposed measure is
that a particular Sasaki-Einstein slice of the Calabi-Yau cone is
singled out by a saddle point calculation. We checked that the
volume of this manifold is properly normalized in field theory
units: we had no additional free parameters in matching the BMN
limits. These volumes are also related to the gravitational
calculation of the conformal anomalies of the field theory.

These matchings show that the conjectured framework can precisely capture
quantitative aspects of strongly coupled theories. The ultimate
objective of this research program is to provide a description of
situations where no other approach seems feasible, such as when
the dual spacetime develops a region of high curvature. However,
before reaching that point, more computations should be done.

It is clear that the calculations that have been done here can be
improved further and one might be able to go beyond BMN limits to
capture more information about string motion in these geometries.
Ideally, one would want to derive that the string motion should
obey the equations of motion associated to a non-linear sigma
model on the corresponding AdS dual geometry.

It is also important to understand more precisely to what extent
the approximations that we have described are applicable, and when
they break down.

\section*{Acknowledgments}

This research was supported in part by the National Science
Foundation under Grant No. PHY05-51164, and by the DOE under grant
DE-FG02-91ER40618.

\appendix

\section{Logarithmic scaling of the Green's function}
\label{sec:log}

In this appendix we show that the Green's function satisfying
\be\label{eq:greensap}
- \nabla^6_6 s(r,r',\q,\q') = - \left(\frac{1}{r^5}
\frac{d}{dr} r^5 \frac{d}{dr} + \frac{1}{r^2} \nabla^2_5 \right)^3
s(r,r',\q,\q') = 64 \pi^3 \delta^{(6)}(r,r',\q,\q') \,,
\ee
has a logarithmic scaling under $r,r' \to \a r, \a r'$, as
advertised in the main text. Recall that $r$ is the radial
direction in the cone (\ref{eq:metric}), whereas the $\q$ are
coordinates on the five dimensional base manifold.

One could find the Green's function using a standard partial wave
expansion for this Laplace-like equation. However, the symmetry $r
\leftrightarrow r'$, crucial for our purposes, may be kept manifest
as follows. Consider the eigenmodes of the related equation
\be\label{eq:eqn2}
- \nabla^2_6 \phi_\l(r,\q) = \l \phi_\l(r,\q) \,.
\ee
These modes give a complete basis of functions. There is an
infinite degeneracy for each value of $\lambda$ given by the modes
\be\label{eq:separate}
\phi_{\l}(r,\q) = \P_{\l,\n}(r) \Q_\n(\theta) \,,
\ee
where
\be
- \nabla^2_5 \Q_\n(\theta) = \n^2 \Q_\n(\theta) \,, \qquad \int
d\q
\sqrt{g_5} \Q^*_\n(\theta) \Q_{\n'}(\theta) = \d_{\n,\n'}
\,.
\ee
The eigenvalues $\nu^2$ are discrete and the lowest is $\nu = 0$,
corresponding to a constant mode on the base of the cone. For each
value of $\nu$, the radial functions are normalised as
\be\label{eq:rnorm}
\int dr r^5 \P_{\l,\n}(r) \P_{\l',\n}(r) = \d(\l-\l') \,.
\ee
The delta function may now be written
\be
\delta^{(6)}(r,r',\q,\q') = \sum_\nu \int_0^\infty d\l \P^*_{\l,\n}(r) \P_{\l,\n}(r') \Q^*_\n(\theta)
\Q_{\n}(\theta') \,.
\ee

Solving the equation (\ref{eq:eqn2}) for the radial part of the
mode (\ref{eq:separate}) and imposing the normalisation
(\ref{eq:rnorm}), one obtains the Bessel function
\be
\P_{\l,\n}(r) = \frac{J_{\sqrt{4+\n^2}}(\sqrt{\l}\, r)}{\sqrt{2} r^2} \,.
\ee
Note that this expression is real. We may now use this expression
to solve for the Green's function in (\ref{eq:greensap}).
Na\"ively, we would like to write the following
\be\label{eq:naive}
s_\text{na\"ive}(r,r',\q,\q') = \sum_{\n}
\int_0^\infty \frac{64 \pi^3 d\l}{\l^3} \P_{\l,\n}(r) \P_{\l,\n}(r') \Q^*_\n(\theta)
\Q_{\n}(\theta')
\,.
\ee
Although this expression formally solves the equation
(\ref{eq:greensap}), it is divergent. The divergence arises as $\l
\to 0$ in the integrand of the $\n = 0$ term. This problem
is entirely expected, due to the fact that the sixth order
equation (\ref{eq:greensap}) has zero modes. Specifically, the six
modes are: $\{1,\log r, r^{\pm 2}, r^{\pm 4}\}$. We can deal with
this divergence as follows.

Firstly, regularize the divergent part of the na\"ive expression
(\ref{eq:naive}):
\be\label{eq:epsilon}
s_\ep = \frac{32 \pi^3}{\text{Vol}(X_5)} \int_\ep^\infty
\frac{d\l J_{2}(\sqrt{\l}\, r) J_{2}(\sqrt{\l}\, r') }{r^2 r'^2 \l^3} +
s_{\n> 0} \,,
\ee
where $s_{\n> 0}$ contains the terms in (\ref{eq:naive}) with $\n
> 0$:
\be\label{eq:nbigger0}
s_{\n > 0} = \sum_{\n > 0} \int_0^\infty
\frac{32 \pi^3 d\l}{r^2 r'^2 \l^3}
J_{\sqrt{4+\n^2}}(\sqrt{\l}\, r) J_{\sqrt{4+\n^2}}(\sqrt{\l}\, r')
\Q^*_\n(\theta) \Q_{\n}(\theta')
\,.
\ee
The integral over $\l$ in this last expression can be performed to
obtain a hypergeometric function. Performing the integral will
break the symmetry $r \leftrightarrow r'$ however, as the result
depends on which of $r$ and $r'$ is bigger.

We can now obtain a finite `renormalised' Green's function via a
minimal substraction
\be\label{eq:result}
s = \lim_{\ep \to 0} \left( s_\ep + \frac{\pi^3 \log \ep}{2
\text{Vol}(X_5)} \right)
\,.
\ee
The expression (\ref{eq:result}) solves the equation for the
Green's function (\ref{eq:greensap}) and is manifestly symmetric
in $r \leftrightarrow r'$. However, we need to check that it is
regular as $r \to 0$ with $r'$ fixed. Taking $r \ll r'$ we obtain
\be
s_\ep(r \ll r') = \frac{4 \pi^3}{\text{Vol}(X_5)}
\int_{\ep\, r'^2}^\infty
\frac{d \l J_2(\sqrt{\l})}{\l^2} + {\mathcal{O}}(r/r') \,,
\ee
which via (\ref{eq:result}) leads to
\be\label{eq:limit}
s(r \ll r') = - \frac{\pi^3 \log r'}{\text{Vol}(X_5)} +
\text{const.} + {\mathcal{O}}(r/r')
\,,
\ee
where the constant is unimportant, as the Green's function is only
defined up to a constant in any case. This expression is
manifestly finite as $r \to 0$.

The expression (\ref{eq:limit}) provides a further nontrivial
check of the result (\ref{eq:result}) as follows. Integrating over
a ball of large radius $r'$
\be
- \int_{B_{r'}} dr d\q \sqrt{g_6}\, \nabla_6^6 s = \pi^3
\left[r^5 \frac{d}{dr} \left( \frac{1}{r^5}
\frac{d}{dr} r^5 \frac{d}{dr} \right)^2 \log r \right]^{r'} = 64 \pi^3 \,,
\ee
as required by (\ref{eq:greensap}).

From (\ref{eq:result}) or (\ref{eq:limit}) it is easy to see that
the Green's function obeys the logarithmic scaling advertised in
(\ref{eq:scaling})
\be
s(\a r,\a r',\q,\q') = s(r,r',\q,\q') - \frac{\pi^3 \log \a}{
\text{Vol}(X_5)} \,.
\ee

\section{Holomorphic polynomials are eigenfunctions at large $N$}

In this appendix we show that wavefunctions of the form
\be
\psi = \psi_0 \Tr P(z) = \sum_i P(z_i) e^{- \sum_j K_j} \,,
\ee
for $P(z)$ a holomorphic polynomial in $z$ with all terms of degree $J$, which in turn is a
holomorphic coordinate on the Calabi-Yau cone with fixed conformal
dimension $c$, are eigenfunctions of the Hamiltonian
(\ref{eq:hamiltonian}) to leading order at large $N$. Some, but
not all, of these arguments essentially appear in appendix A of
\cite{Berenstein:2007wz}. These arguments go through if $P$ is not
holomorphic, but simply a harmonic function on the Calabi-Yau
cone.

Holomorphy implies $\nabla^2 P(z) = 0$ and scaling dimension $c$
of $z$ implies $r \pa_r z = c z$. Straightforward algebra then
shows that
\be\label{eq:interm}
H \psi = (E_0 + c J) \psi + \psi_0 \sum_i \nabla_i P(z_i) \cdot
\sum_{j \neq i} \nabla_i s(z_i,z_j) \,.
\ee
We now show that the last term vanishes to leading order at large
$N$.

In the continuum large $N$ limit, the last term in
(\ref{eq:interm}) is proportional to
\be\label{eq:interm2}
\int d^6x \rho(x) \nabla_x P(z_x) \cdot \int d^6y \rho(y) \nabla_x
s(z_x,z_y)\,.
\ee
If $P(z)$ is a polynomial with not too high degree, as in the case
$P(z) = z^J$ we considered in section 8 above, then the backreaction of $P(z)$
onto the eigenvalue distribution is subleading at large $N$.
Therefore in (\ref{eq:interm2}) we may take the $\rho(x)$ to be
the ground state (\ref{eq:solution}). In particular, this
distribution adds no extra dependence on the coordinates $\theta$
of $X_5$. The integral over $d^6y$ includes an integral over
$X_5$. From the fact that $\int d\q
\sqrt{g_5} \Theta_\n(\q) = 0$ for $\nu > 0$ and from the
expression (\ref{eq:nbigger0}), only the part of $s(z_x,z_y)$ that
is independent of both $\theta_x$ and $\theta_y$ survives the
$\theta_y$ integral. Thus (\ref{eq:interm2}) is proportional to
\be\label{eq:interm3}
\int d^5\q \sqrt{g_5} \pa_r P(z) \,.
\ee
The final step is now to show that $P(z)$, and hence also $\pa_r
P(z)$, is a nontrivial eigenfunction of the Laplacian on $X_5$,
and therefore the integral (\ref{eq:interm3}) vanishes. From
holomorphy we have
\be\label{eq:laplacian}
\nabla^2 P(z) = \left(\frac{1}{r^5}
\frac{d}{dr} r^5 \frac{d}{dr} + \frac{1}{r^2} \nabla^2_5 \right) P(z) =
0\,.
\ee
The scaling dimension of $z$ implies that each monomial in $P(z)$ is of
the form $P_J(z) = r^{c J} F_J(\theta)$. It is immediately seen that
(\ref{eq:laplacian}) implies that
\be
- \nabla^2_5 F_J(\theta) = c J(c J+4) F_J(\theta) \,.
\ee
Therefore $P_J(z)$ is a harmonic of the Laplacian on $X_5$, as we
required.

The upshot of the preceding paragraph is that (\ref{eq:interm2})
does indeed vanish and hence the holomorphic polynomial does give
an eigenfunction, as claimed.

\section{Coherent states and orthogonality}

In the large $N$ limit, single trace operators are supposed to be
related to single string states. To the extent that these are
free, one can build coherent states of these traces. Formally, we
would want to consider a coherent state as an exponential of a
raising operator. In our identification, we have said that $\Tr
h(x)$ is a single graviton state, so a coherent state of gravitons
would be described formally by
\begin{equation}
\psi_\text{coh} \sim e^{ \alpha \Tr h(x)} \psi_0 \,.
\end{equation}
We can try to understand the distribution of particles on the cone
that is associated to this wave function. We do this by thinking
of $\alpha$ as a formal parameter (usually $h(x)$ will grow faster
at infinity than the decay of $\psi_0$, which is just gaussian
decay).

If we replace $\Tr h(x)$ by $\int \rho(x) h(x)$, as is required
for the large $N$ limit, we can repeat the arguments made in
studying equation (\ref{eq:lagrange}) to show that once again the
dominant semiclassical density of eigenvalues is a singular
distribution. It was suggested in \cite{Berenstein:2005aa} that
having singular distributions of particles in the saddle point
limit is exactly the type of situation that leads to classical
gravity solutions. This supports the proposal made above for the
wave functions associated to non-BPS gravitons.

Unfortunately, it seems that the corresponding wave functions are
not eigenfunctions of the full effective Schr\"odinger equation
with the measure added. This has already been seen for the case of
${\cal N}=4 $ SYM \cite{Berenstein:2007wz}. There is no new effect
that shows up in this more general case that is not there in the
case of maximal supersymmetry. Moreover, as shown in appendix B,
they become eigenstates to leading order in the large $N$ limit.

One can also show that these  single trace wave functions
(including the measure) are approximately orthogonal to the ground
state and to each other. One would need to evaluate the overlap
\begin{equation}
\int e^{-r^2} \mu^2 \Tr h_1(x) \Tr h_2(x) \,.
\end{equation}
The idea to show approximate orthogonality is that the overlap is
dominated by the saddle of the ground state. Then the dependence
of $\mu^2$ on eigenvalue $i$, that is written as
\begin{equation}
\mu^2_i= \exp(- \int \rho(x) s(x_i,x)) \,,
\end{equation}
can be approximated by a function that depends only on the radial
variable $r_i$, but not on the angular variables. Using the
product decomposition of $h$ into a radial and angular part, we
see that the orthogonality of the angular part of the wave
functions makes these single trace perturbations orthogonal to
each other (unless $h_1\sim h_2^*$).

For ${\cal N}=4 $ SYM this statement is exact and follows from
orthogonality of different unitary representations of the $SO(6)$
symmetry group. This orthogonality is also exact in ${\cal N}=1$
cases when $h_1$ and $h_2$ have different R-charges. These
arguments can be extended further, and suggest that the standard
large $N$ counting arguments are applicable in some generality.

The arguments show that the details of the calculations depend on
various properties of harmonic analysis on the Saski-Einstein
manifold and the particular saddle point we found that determines
the vacuum structure. To our knowledge, the most general study of
the spectrum of the scalar Laplacian has been done in \cite{Oota}.

\end{document}